\documentstyle[12pt,aasms4,epsfig,psfig,lscape]{article}

\newcommand{\3}{3C~279}
\newcommand{\sax}{\emph{Beppo}SAX }
\newcommand{\bsax}{\emph{\textbf{Beppo}}SAX}

\begin{document}

\title{Spectral Energy Distributions of 3C~279 revisited: \bsax\,observations and variability models}

\author{ L.~Ballo$^1$, L.~Maraschi$^1$, F.~Tavecchio$^1$, A.~Celotti$^2$, G.~Fossati$^3$, G.~Ghisellini$^4$, E.~Pian$^5$, C.~M.~Raiteri$^6$, G.~Tagliaferri$^4$, A.~Treves$^7$, C.~M.~Urry$^8$, M.~Villata$^6$}

\altaffiltext{1}{Osservatorio Astronomico di Brera, via Brera 28, 20121 Milano, Italy}
\altaffiltext{2}{Scuola Internazionale Superiore di Studi Avanzati/International School for Advanced Studies, Astrophysics Sector, via Beirut 4, 34014 Trieste, Italy}
\altaffiltext{3}{Center for Astrophysics and Space Sciences, University of California, San Diego, 9500 Gilman Drive, Code 0424, La Jolla, CA 92093, USA }
\altaffiltext{4}{Osservatorio Astronomico di Brera, via Bianchi 46, 23807 Merate, Italy}
\altaffiltext{5}{Osservatorio Astronomico di Trieste, via G. B. Tiepolo 11, 34131 Trieste, Italy}
\altaffiltext{6}{Osservatorio Astronomico di Torino, Strada Osservatorio 20, 10025 Pino Torinese (TO), Italy}
\altaffiltext{7}{Dipartimento di Fisica, Universit\`a dell'Insubria, via Valleggio 11, 22100 Como, Italy}
\altaffiltext{8}{Yale Center for Astronomy and Astrophysics, Yale University, New Haven, CT 208121, USA}

\setcounter{footnote}{0}

\begin{abstract}

We present the \sax  observations of \3 performed in 1997~January 13~-~23 simultaneously with $\gamma$-ray pointings with EGRET on board CGRO and optical observations at the Torino Observatory (1997~January 11~-~18).
ISO data close to this epoch are also available.
We compare the derived Spectral Energy Distribution with those obtained at all other epochs with adequate multiwavelength coverage.
Simple spectral models fitted to the multiepoch SEDs suggest that changes in intensity can be ascribed mainly to the variation of the bulk Lorentz factor of the plasma in the jet.

\end{abstract}


\section{Introduction}

In the last ten years about 60 blazars (BL~Lac objects and flat-spectrum radio quasars) have been detected in $\gamma$-rays by EGRET (Mukherjee~et~al.~1997, Hartman~et~al.~1999).
The observations show that a large fraction of the total power is emitted in this band, and many of these objects exhibit rapid  variability (on timescales of hours or days) at the highest energies.
The high  $\gamma$-ray luminosities and short variability timescales impose strong and general constraints on the emitting region; in particular, independent of other evidence, the $\gamma$-ray transparency condition implies that the emitting plasma moves  relativistically.

The mechanism of $\gamma$-ray production is most likely inverse Compton scattering: the seed photons could be produced both within the jet via synchrotron radiation, which is responsible for the emission from radio to UV, (Synchrotron Self Compton~[SSC], see Maraschi,~Ghisellini~\&~Celotti~1992), and outside the jet (External Compton~[EC]) by the accretion disk (Dermer~\&~Schlickeiser~1993) or by the broad line region clouds (Sikora,~Begelman~\&~Rees~1994).
The relative weight of the two sources of seed photons is given by the ratio of their respective energy densities as seen in the frame of the emitting matter (Sikora,~Begelman~\&~Rees~1994).
In flat spectrum radio quasars, since the external photons typically have much higher frequencies than synchrotron photons, when both processes SSC and EC are important the former dominates the SED at low energies (in the X-ray band, see~Kubo~et~al.~1998), the latter in the $\gamma$-ray range; significant information about the internal structure of the jet can thus be obtained from observations covering the full X-ray to $\gamma$-ray spectral region.

Another open problem is the physical cause of the variability.
Flares could be due to propagating shocks causing transient acceleration of relativistic electrons (see e.g.~Marscher~\&~Gear~1985).
Recently an idea initially proposed by Rees~(1978) was worked out and applied in particular to \3 by Spada~et~al.~(2001)(see also Ghisellini~1999): briefly, the collision of shock fronts travelling in the jet with different bulk Lorentz factors can yield efficient impulsive particle acceleration leading to enhanced radiation (Spada~et~al.~2001; Ghisellini~1999).

In any case both the synchrotron and inverse Compton emission from a single electron population are expected to vary in a correlated fashion, but for a given change in the electron spectrum the amplitudes of the SSC and EC variations will be different (Ghisellini~\&~Maraschi~1996, Sikora~1997).
Furthermore inhomogeneities (clouds) in the medium close to the jet may act as mirrors for the synchrotron photons (Ghisellini~\&~Madau~1996, Boettcher~\&~Dermer~1998, Bednarek~1998). 
Variability may also result from changes in the beaming factor, due to geometric effects (see~Villata~\&~Raiteri~1999).
All of these hypotheses predict a particular relative variability in the synchrotron and inverse Compton components, so a study of simultaneous variations in different bands can provide important information about the nature of phenomena occurring in blazar jets.

\3 ($z=0.538$) was the first blazar discovered as a $\gamma$-ray source with the \emph{Compton Gamma Ray Observatory} (CGRO; Hartman~et~al.~1992) and one of the brightest
in this band, and as such was the target of several multifrequency campaigns (for the most recent re-analysis of all the data see Hartman~et~al.~2001, hereafter~H01). 
It was chosen for observations with \sax (1997~January 13~-~23) simultaneous with an EGRET pointing and optical observations (1997~January 11~-~18)
to improve our knowledge of the detailed spectral shape from X-rays to $\gamma$-rays
and to continue the study of its variability at high energies.
Since \sax covers a very wide energy range ($0.1-100\,$keV), 
it is ideal to study the connection between the X and $\gamma$-ray continuum
where the transition from the SSC to the EC process should take place.
In the present work a complete analysis of the \sax observations and associated SED are fully presented, combined with a discussion of the historical SEDs of 3C~279 with best multiwavelength coverage.
Preliminary results were published in Maraschi~et~al.~(1998), Maraschi~(2000), Maraschi~et~al.~(2000), Maraschi~\&~Tavecchio~(2001a) and Maraschi~\&~Tavecchio~(2001b).
The \sax data are also included in H01.

The structure of the paper is the following: in \S2 we present the analysis and results
 of the \sax observations of 1997; in \S3 we construct the overall SED of 1997 
and  compare it with other emission states for which simultaneous or 
quasi-simultaneous multifrequency data are available; in \S4 we apply a
simple spectral model including both the SSC and EC processes to reproduce the
observed SEDs at different epochs and derive
 physical parameters for the emitting plasma.
Conclusions are presented in \S5.
Throughout the paper we assume $H_0=65\,$km~s$^{-1}\,$Mpc$^{-1}\,$ and $q_0=0.5$.

\section{\bsax\,observations of 1997}

The scientific payload of the Italian-Dutch X-ray satellite \emph{Beppo}SAX\footnote{\small http://www.sdc.asi.it} (see Boella~et~al.~1997) consists of four coaligned Narrow Field Instruments (NFIs) and two Wide Field Cameras.
Two of the NFIs use concentrators to focalize X-rays: the Low Energy Concentrator Spectrometer (LECS; one unit) has a detector sensitive to soft-medium X-ray photons (0.1-10 keV), while the Medium Concentrator Spectrometer (MECS; three units) detects photons in the energy range 1.3-10 keV.
The Phoswich Detector System (PDS), sensitive from 15 to 300 keV nominally, consisting of four identical units, uses rocking collimators so as to monitor source and background simultaneously with interchangeable units.
We will not consider here data from the fourth NFI, a High Pressure Gas Scintillation Proportional Counter (HPGSPC).

The observations of \3 were planned to be simultaneous with a continuous pointing at the source with EGRET and with optical observations performed at the Torino Observatory.
They consisted of 5 pointings, distributed over 10 days (1997~January 13~-~23) to cover the typical two-week duration of an EGRET observation.
Exposure times and observed count rates in the various detectors are reported in the \sax journal of observations (see Table~\ref{tab:saxlog}).
We searched for variability both within each observation and between the different observations using the $\chi^2\,$test against constancy: no significant flux variations were detected; we therefore combined all the observations deriving a cumulative spectrum.

The \sax spectral data were analyzed using the standard software packages XSELECT (v1.4) and XSPEC (v10.0) and the September 97 version of the calibration files released by the \sax Scientific Data Center (SDC).
From the event files we extracted the LECS and MECS spectra in circular regions centered around the source with radii of 8$^{\prime }$ and 6$^{\prime }$ respectively (see the SAX Analysis Cookbook\footnote{\small ftp://www.sdc.asi.it/pub/sax/doc/software\_docs/saxabc\_v1.2.ps.gz}).
The PDS spectra extracted with the standard pipeline with the rise-time correction were provided directly by the \sax SDC.

For the spectral analysis we considered the LECS data in the restricted energy range 0.1-4 keV, because of unsolved calibration problems at higher energies.
PDS data above $180\,$keV were discarded since the count rate was too close to the instrumental detection limit to be reliable.
Background spectra extracted from blank field observations at the same detector position as the source were used for background subtraction in the LECS and MECS, while for the PDS the simultaneously measured off-source background was used.
We fitted rebinned LECS, MECS and PDS net spectra jointly, allowing for two normalization factors to take into account uncertainties in the intercalibration of different instruments (see SAX Cookbook).

The total LECS+MECS+PDS spectrum is well described by a single power-law model, with an absorption column consistent with the galactic value ($N_H=(2.21\pm0.1)\cdot10^{20}\,$cm$^{-2}$, as reported in Elvis,~Wilkes~\&~Lockman~1989); the results of the spectral fits are summarized in Table~\ref{tab:saxfit}.
The residuals of the fit show a weak excess around $E\simeq0.5\,$keV in the LECS data (see Fig.~[\ref{fig:fitpolaw}], upper panel): we modelled the whole spectrum with either a broken power law or a power law with a black body component.
In both cases the excess observed at low energy is partly accounted for (see Fig.~[\ref{fig:fitpolaw}], lower panel), but requires a column density higher than the galactic value.
In any case the reduction in $\chi^2$ is not significant and we considered only the power law model in the rest of our analysis.

All through the analysis the LECS/MECS normalization ratio was allowed to vary; the best fit value obtained is $0.85$, in agreement with the intervals proposed by SAX Cookbook.
We fixed the intercalibration between PDS and MECS to the usual value, $0.85$: the fit with a free normalization factor gives a slightly higher value (even if in the acceptable range), without considerable improvement in the $\chi^2$.

\section{Spectral Energy Distributions and models}

\subsection{1997}

Simultaneously to the \sax observations \3 was monitored in $\gamma$-rays by EGRET and in the optical at the Torino Observatory.
The $\gamma$-ray data are reported in H01.

R, V and B band magnitudes were obtained between 1997~January 11~-~18 at the $1.05\,$meter REOSC telescope, equipped with a $1296\times1152$ pixels Astromed CCD camera with a $0.467\,$arcsec/pixel image scale.
Data were reduced with the ROBIN procedure locally developped, which includes bias subtraction, flat field correction, and circular Gaussian fit.
Magnitudes are reported in Table~\ref{tab:optlog}.
A search in the UMRAO database provided radio data at $4.8\,$GHz (1997~January 21~-~22), $8.0\,$GHz (1997~January~24) and $14.5\,$GHz (1997~January~14).

In 1996~December~19 (only one month before the \sax pointings), \3 was observed by ISOPHOT, the photometer on board ISO, as part of a program involving numerous AGN.
The source was detected in seven bands, from $4.8\,\mu$m to $170\,\mu$m; results are reported in Haas~et~al.~(1998).

All the above fluxes together with the deconvolved \sax $0.1-100\,$keV spectrum are shown in Fig.~[\ref{fig:sed97}].
The X-ray spectrum is flat, while the connection with the EGRET $\gamma$-ray data indicates a steepening of the SED between the two ranges.
The $\gamma$-ray flux is slightly higher than that measured in 1993, when the source was in a very low emission state (see the next section); the same occurs for the optical luminosity.

Finally, it is notable that the far infrared peak observed by ISO dominates the bolometric luminosity.
Integrating the measured fluxes, Haas~et~al.~(1998) find $L_{IR}=2\cdot10^{47}\,$erg~s$^{-1}$.
This value is similar to that found for three other flat spectrum radio quasars examined by the same authors; however, while in the other FSRQs the infrared emission is a smooth continuation of the synchrotron spectrum, in the case of \3 the IR spectrum shows a narrow peak suggestive of a thermal origin.

However previous observations with IRAS in four bands from $12\,\mu$m to $100\,\mu$m (see~Impey~\&~Neugebauer~1988, Moshir~1990) yielded a $60\,\mu$m flux density and three upper limits lower than the fluxes measured by ISO, principally at $60\,\mu$m (a factor of $\sim9$).
If this discrepancy is real (see Haas~et~al.~1998) the implied variability would not be consistent with thermal emission from a region of about $1\,$kpc.

Moreover from the shape of the ultraviolet spectrum in the lowest state one can derive an upper limit of $\sim2\cdot10^{45}\,$erg~s$^{-1}$ to the UV luminosity of a possible accretion disk in \3 (Pian~et~al.~1999).
Since in the case of \3 the line of sight is at small angle to the axis of the relativistic jet which is presumably close to that of the putative accretion disk, it is difficult to suppose the existence of a hidden nuclear UV component with intensity similar to that of the IR emission.
Thus it is not easy to understand how the dust could be heated unless the AGN underwent a long phase of higher activity in the past or a highly luminous starburst is taking place in this galaxy.

\subsection{Multiepoch SEDs}

One of the characteristic properties of blazars is the extreme variability, observed at all wavelengths but much accentuated at high energies.
Understanding this trait can provide important information about mechanisms operating in the inner regions of the source.
Therefore, after collecting data provided from the campaign organized in 1997, we revisited several SEDs of \3 obtained with simultaneous or quasi-simultaneous multiwavelength data, with the scope of comparing different states of emission.
In the following we briefly recapitulate basic information on the SEDs used: the first refers to the epoch of the $\gamma$-ray discovery of \3 in 1991, the other three are based on organized campaigns with wide multiwavelength coverage (1993 and 1996~pre-flare, flare).

\begin{description}

 \item[1991:] The first detection of $\gamma$-ray emission dates from 1991 June 16~-~28: the flux measured by EGRET was very intense, with  evidence of a flare on a timescale of days (Kniffen~et~al.~1993); a second observation (1991 October 3~-~17) revealed a lower flux level.
A literature search provided quasi-simultaneous data to the EGRET observations (see Hartman~et~al.~1996; Bonnell,~Vestrand~\&~Stacy~1994) to assemble the SED shown in Fig.~[\ref{fig:sed9193}] (upper panel).
IR data were obtained the 1991 June 16 and the optical one is an average over the epoch of the $\gamma$-ray flare, while the UV observations date from 1991 July 27, without resorting to the correlation between UV and R-band fluxes applied in Hartman~et~al.~(1996).

 \item [1993:] This first multiwavelength campaign organized after the discovery found \3 in a low state: the emission had faded dramatically at all frequencies above $10^{14}\,$Hz, while the flux variations at radio to millimeter wavelengths were minor.
To assemble the SED in Fig.~[\ref{fig:sed9193}] (lower panel) we used data from Maraschi~et~al.~(1994), but differently from that work the EGRET observation (having a $4\,\sigma$ level) is taken as an upper limit.

 \item [1996:] Throughout this second multiwavelength campaign \3 was in a high state, similar to that of 1991; during the observation there was an extraordinary flare lasting 2-3 days, particularly intense in the $\gamma$-ray band.
Following Wehrle~et~al.~(1998), we divided the light curve in two parts, averaging the observed fluxes in the period January 24~-~28, preceding the flare, and  in a 2-day window centered on the $\gamma$-ray peak. 
The resulting SEDs are shown in Fig.~[\ref{fig:sed96}].

\end{description}

\section{Interpretation}

In order to reproduce the observed SEDs we used a model with minimal assumptions.
The emitting region is described as a homogeneous sphere with radius $R$ endowed with a magnetic field $B$ and moving with bulk Lorentz factor $\Gamma$; the line of sight is at an angle $\theta$ with respect to the jet axis, so that the  Doppler factor $\delta=1/[\Gamma(1-\beta\cos\theta)]$ is relevant for relating observed quantities with rest frame quantities.
The source is filled with relativistic electrons, whose energy spectrum is described by a broken power law between $\gamma_{min}$ and $\gamma_{max}$, which is determined by four parameters, the break energy $\gamma_b(\gg1)$, the normalization $k$ and the spectral indices of the asymptotic power laws below and above the break, respectively $n_1<3$ and $n_2>3$ (see~Tavecchio,~Maraschi~\&~Ghisellini~1998):
\begin{displaymath}
 N(\gamma)=k\gamma^{-n_1}\left(1+\frac{\gamma}{\gamma_b}\right)^{n_1-n_2}
\end{displaymath}
This form for the distribution function has been assumed in order to describe the curved shape of the SED.
We used $\gamma_{max}=5\cdot10^{4}$ for all the states.
The value of the minimum energy of the scattering electrons is uncertain.
We assumed $\gamma_{min}=1$ as inferred in the case of other emission line blazars from the lack of spectral breaks in the soft X-ray continuum (see Tavecchio~et~al.~2000).

Electrons emit via the synchrotron and inverse Compton mechanisms.
At low frequencies the synchrotron spectra are limited by the self-absorption frequency.
Below that frequency the model has the standard self-absorbed spectrum with slope $5/2$.
Additional emission components from regions farther out in the jet are necessary to account for the spectra at lower frequencies, as indeed expected if the flat radio spectra of blazars are due to the superposition of different self-absorbed components from different locations in the jet.
The SEDs calculated here refer to the innermost emitting region.

The inverse Compton spectra have been calculated using the full Klein-Nishina cross-section (Jones~1968; see also Blumenthal~\&~Gould~1970) and taking into account the beaming of the external radiation in the blob frame as pointed out in Dermer~(1995).
There the author uses a single power law as electron distribution and a monochromatic external radiation.
In our case the flux calculation can be made only numerically because we use a more complex electron distribution.

We neglected the direct contribution of photons from the accretion disk since it is strongly red-shifted in the comoving frame when the bulk Lorentz factor is large and the $\gamma$-ray emission occurs at distance $\geq10^{17}\,$cm (Sikora,~Begelman~\&~Rees~1994).
The dominant contribution to the external radiation can be identified with that produced by the BLR via reflection/reprocessing of the disk emission.
The spectrum of the BLR is described by a black body peaking at a frequency $\nu_{ext}=10^{15}\,$Hz with a luminosity $L_{blr}$ diluted in a spherical region with radius $R_{blr}$.
For \3 the BLR luminosity was estimated in Celotti,~Padovani~\&~Ghisellini~(1997) from the observed emission lines (see also Francis~et~al.~1991) as $L_{blr}=6\cdot10^{44}$erg~s$^{-1}$.
This value corresponds to $\sim30\%$ of the accretion disk luminosity estimated by Pian~et~al.~(1999) from the UV observations during the 1993 campaign.
Applying the model to the 1993 state, the one with lowest emission, we found that the best reproduction was obtained for an energy density $u_{blr}\sim10^{-4}\,$erg~cm$^{-3}$; this corresponds to a BLR radius $R_{blr}=4\cdot10^{18}\,$cm, consistent with the value obtained from the $R_{blr}-L_{blr}$ relation found by Kaspi~et~al.~(2000).
This value was then kept fixed for all states.

We also estimated the possible contribution of the observed IR radiation.
From the luminosity and temperature given by Haas~et~al.~(1998) we derived the minimum radius of the emitting region, $R_{min}=\sqrt{\frac{L_{IR}}{4\,\pi\sigma\,T^4_{IR}}}\sim10^{20}-10^{21}\,$cm, and a maximum energy density $u_{IR}\sim10^{-8}-10^{-5}\,$erg~cm$^{-3}$, negligible with respect to the BLR energy density calculated above.

Finally, modelling the emission from blazars it is common to adopt the relation $\delta\sim\Gamma$, corresponding to an angle of view $\theta=1/\Gamma$.
However studying different states of the same object we intended to allow variations of $\Gamma$ and $\delta$ but not of the angle of sight; so we fixed it to a value $\theta=3^{\circ}$, leaving the Lorentz factor $\Gamma$ free to vary and computing the corresponding value of $\delta$.

The computed models are shown together with the data 
in Figs.~[\ref{fig:sed97}], [\ref{fig:sed9193}] and~[\ref{fig:sed96}] with
the different components (synchrotron,  disk radiation, SSC and EC) 
as well as their sum plotted separately.
In Fig.~[\ref{fig:cfr}] the models 
(sum of all the contributions) reproducing the  SEDs of the different
states are compared with each other.

We tried to reproduce the different levels varying the smallest possible 
number of parameters.
Our models are consistent with the observed variations being essentially 
due to the bulk Lorentz factor $\Gamma$ (also in the short timescale 
flare observed in~1996 ); the parameters required are given 
in Table~\ref{tab:parameters}.
The goodness of the choice of a $\gamma_{min}=1$ was verified a posteriori: considering all the states we found that if $\gamma_{min}>5$ the models cannot reproduce the hard X-ray/soft $\gamma$-ray data.
The magnetic field is $\simeq0.5\,$G in all states, while the size of emitting region and the Doppler factor used allow variability timescales $t_{var}=\frac{R}{\delta\,c}$ from $1$ to $2$ days, consistent with the observations.
We note that the small angle of view ($\theta=3^{\circ}$) and the high bulk Lorentz factor (from $6$ to $17$) used involve relatively high superluminal speeds, from $\beta_{app}\sim3-4$ in the low states to $\beta_{app}\sim12-15$ in the high levels.
Superluminal velocities observed in \3 range from $3c-5c$ (Unwin~et.~al.~1989; Carrara~et~al.~1993) to $4.8c-7.5c$ derived from a long-term high-frequency VLBI monitoring of six superluminal components in the relativistic jet in \3 (see Wehrle~et~al.~2001).
Higher velocities have also been measured in the past: Cotton~et.~al.~(1979) originally found a value of $15c$ (speeds have been expressed assuming $H_0=70\,$km~s$^{-1}\,$Mpc$^{-1}$ and $q_0=0.1$).
These velocities refer however to regions further away along the jet, 
while our data concern the emission from much smaller scales. Thus the
comparison, though interesting, can only be indicative. We would like to mention that in the
internal shock model it is predicted that shells with high $\Gamma$
will slow down to an average value of $\Gamma$ further out.

In the models above  the X-ray emission is attributed to the SSC mechanism,
 while the $\gamma$-ray emission is ascribed to the EC process; 
in this case, supposing a variation in the bulk Lorentz 
factor $\Gamma$, one expects that the relative change of the peak emission
 of these two components follows the relation 
$F_{EC}\propto F_{SSC}^{3/2}$ (see Ghisellini~\&~Maraschi~1996).
We searched for a correlation between X-ray and $\gamma$-ray fluxes 
in 3C~279 using all the available simultaneous measurements in 
the two bands, derived from H01; 
in Fig.~[\ref{fig:correl}] we plot $\nu_{\gamma}F_{\gamma}\,$~vs~$\,\nu_XF_X$.
The X-ray data are not completely homogeneous.
When possible, we used the flux measured at $3\,$keV. Otherwise
 we took the available flux 
 at a slightly different energy ($\sim2\,$keV).
For the EGRET data $\nu_{\gamma}$ is taken at $167\,$MeV 
(frequency at which observations for all the states are given 
in H01).
In some cases H01 gives two fluxes corresponding
to energies of $121\,$MeV and $208\,$MeV respectively.
In these cases we used an average flux.
For all the measurements the plotted errors are the maximum between 
the value provided and the $5\%$ of the flux, 
in order to take into account intercalibration problems.
The straight line, obtained with a fit that considered the errors 
in both ranges, has slope $1.49\pm0.13$, in good agreement with the 
expectation.

\section{Discussion and Conclusions}

We have analyzed the X-ray spectra of \3 obtained from the \sax satellite 
in 1997 January. The result of our analysis is that the featureless X-ray
continuum in the $2-100\,$keV energy band is well represented by a
power-law with $\alpha\sim0.66$.  The observations from \sax were part of
an organized campaign involving also $\gamma$-ray and optical
measurements; adding radio data from the UMRAO database and
IR data from ISO, 1 month earlier than the high energy observations,    
 a quasi-simultaneous SED from radio to $\gamma$-rays was derived.

We also revisited the overall SEDs of four other
different states of \3 with wide simultaneous spectral coverage, 
from 1991 to 1997.
We modelled the observed SEDs with the widely used
homogeneous synchrotron-Inverse Compton model, estimating the physical
parameters in the emission region. This was done using the minimum 
possible number of free parameters.

The model used in this work (simpler than the one adopted in H01) is fully specified by eight parameters: the four parameters of the electron distribution, the magnetic field, the size of the emitting region, the bulk Lorentz factor and the angle of view.
In addition the external radiation energy density and the typical frequency of the photons enter in the model.
The latter is practically fixed at $\nu_{ext}\simeq10^{15}\,$Hz, while the former has been inferred estimating the luminosity and the dimension of the BLR: so they can't be considered free parameters.
From an observed SED we can in principle obtain six quantities, namely the synchrotron peak frequency and luminosity, the inverse Compton peak frequency and luminosity and the spectral indices of the synchrotron component after the peak and that of the inverse Compton component before the peak (connected to the indices of the electron distribution $n_2$ and $n_1$ respectively).
In addition to these six quantities the typical variability timescale can give an upper limit to the size of the source.
Fixing the angle of view, the total number of observational constraints is equal to the number of parameters of the model which can therefore in principle be fully constrained (see Tavecchio~et~al.~2000).
In the case of \3 the main observational uncertainty is the position of synchrotron peak, falling in the poorly covered FIR band.
However given the fact that we consider $5$ states, we feel that within the model assumptions our parameter determination is robust.

We confirm that the bulk of the $\gamma$-ray (MeV-GeV) emission
can be modelled as IC radiation produced by electrons in the jet
upscattering soft ambient photons reprocessed in the Broad Line
Region as originally proposed by Sikora,~Begelman~\&~Rees~(1994) (for a recent review of blazar models, see Sikora~\&~Madejski~2001)
while the X-ray continuum should be due
to Synchrotron-Self Compton radiation.
Notably the relation between the 
$\gamma$-ray and X-ray fluxes of all the available simultaneous observations
follows the relation expected under this hypothesis.

The most interesting result of our study is that the SEDs 
 of the 5 different states considered here
can be reproduced varying {\it in a substantial way} only one parameter,
namely the bulk Lorentz factor $\Gamma$ of the emitting region, and
considering all the other physical quantities (size, magnetic field,
electron distribution) as almost constant.  
A significant variation of $\Gamma$ can be understood if the jet is
energized by the injection of shells with initially different values of
the bulk Lorentz factor as  proposed
in the internal shock model (see~Spada~et~al.~2001; Ghisellini~1999).
 The collision between successive shells naturally predicts the production 
of a new shell with an intermediate $\Gamma$ on short timescales 
as necessary in the case of the 1996~flare.

Other causes of variability may coexist with the above as it is well
possible to reproduce the observed variability varying {\it more} than one
parameter. Even then however, it is not possible to reproduce the 
variability behaviour {\it without} varying $\Gamma$ (H01).
In particular we cannot exclude that the peaks of the 
main emission components shift to higher energies in brighter
states. This is in fact suggested by the hardening of the $\gamma$-ray
spectrum. However the position of the synchrotron
peak  in this source as well as in many emission line blazars is 
difficult to constrain since it falls in the FIR band where observations
are difficult. 

Even taking into account the above uncertainties it seems that
the  variability behaviour of \3 differs form that of Mkn 501,
one of the best studied and most ``extreme'' or ``high energy peaked'' 
BL Lac objects in the spectral sequence discussed by Fossati~et~al.~(1998) 
(see Costamante~et~al.~2001).
In the latter case the large spectral variability observed in X-rays 
on short and long timescales, causing a substantial shift in the peak 
of the synchrotron emission, appears to be mainly due to the variation
of the energy of electrons emitting at the peak (those at
$\gamma _b$, Tavecchio~et~al.~2001) presumably due to a change in the
acceleration process {\it not associated with a significant change in
$\Gamma$}. In fact the emission of Mkn 501 in the soft X-ray band varies
much less than in the medium-hard band implying strong intrinsic
spectral changes and small if any variations in $\Gamma$, at least
within a homogeneous model.

These issues are clearly important for a physical understanding of the
 variability of blazars which could be driven by
somewhat different processes in different ``types'' of blazars (emission line
blazars~vs.~BL~Lacs or high luminosity~vs.~low luminosity, see Ghisellini~et~al.~1998, Ghisellini~\&~Celotti~2001).
Simultaneous access to X-ray, FIR and $\gamma$-ray plus ground based facilities 
as will be possible in the next years will allow us to test new ideas in this field.

\clearpage

\acknowledgments{We thank the $Beppo$SAX SDC for providings us with the cleaned data.
This research has made use of data from the University of Michigan Radio Astronomy Observatory which is supported by funds from the University of Michigan.
We used also the NASA/IPAC Extragalactic Database (NED), which is operated by the Jet Propulsion Laboratory, California Institute of Technology, under contract with the National Aeronautics and Space Administration.
This work was partly supported by the Italian Ministry for University and Research (MURST) under grant~Cofin2000 and by the Italian Space Agency (ASI) under grant~I-R-105-00.

\clearpage

\vskip 1.5 truecm

\centerline{ \bf Figure Captions}

\vskip 1 truecm

\figcaption[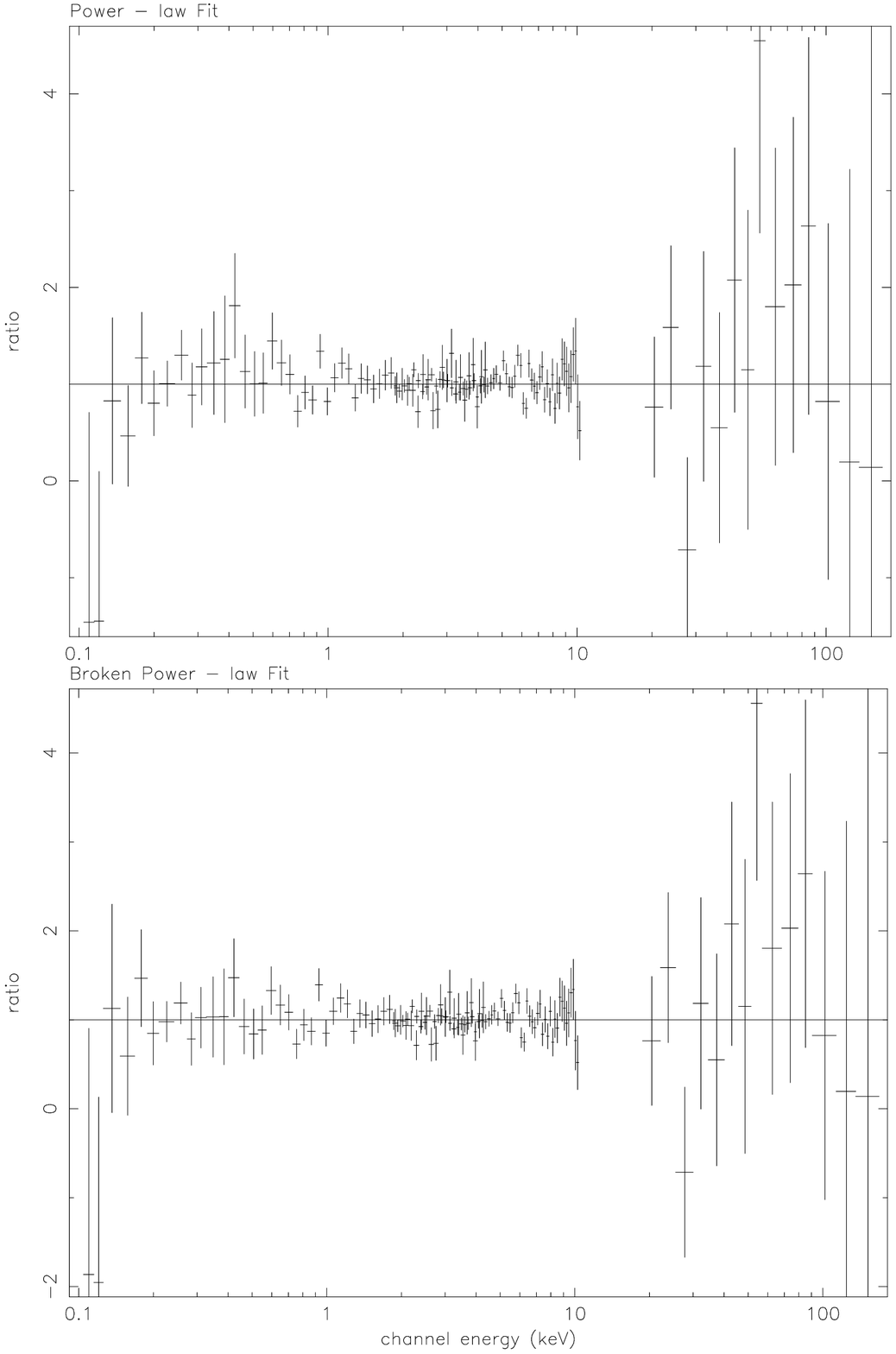]{Residuals of the \sax data fitted with a power law (upper panel) and a broken power law (lower panel); the absorption is free. Note that the weak excess observed at low energy in the first case is partly accounted for in the second one.\label{fig:fitpolaw} }

\figcaption[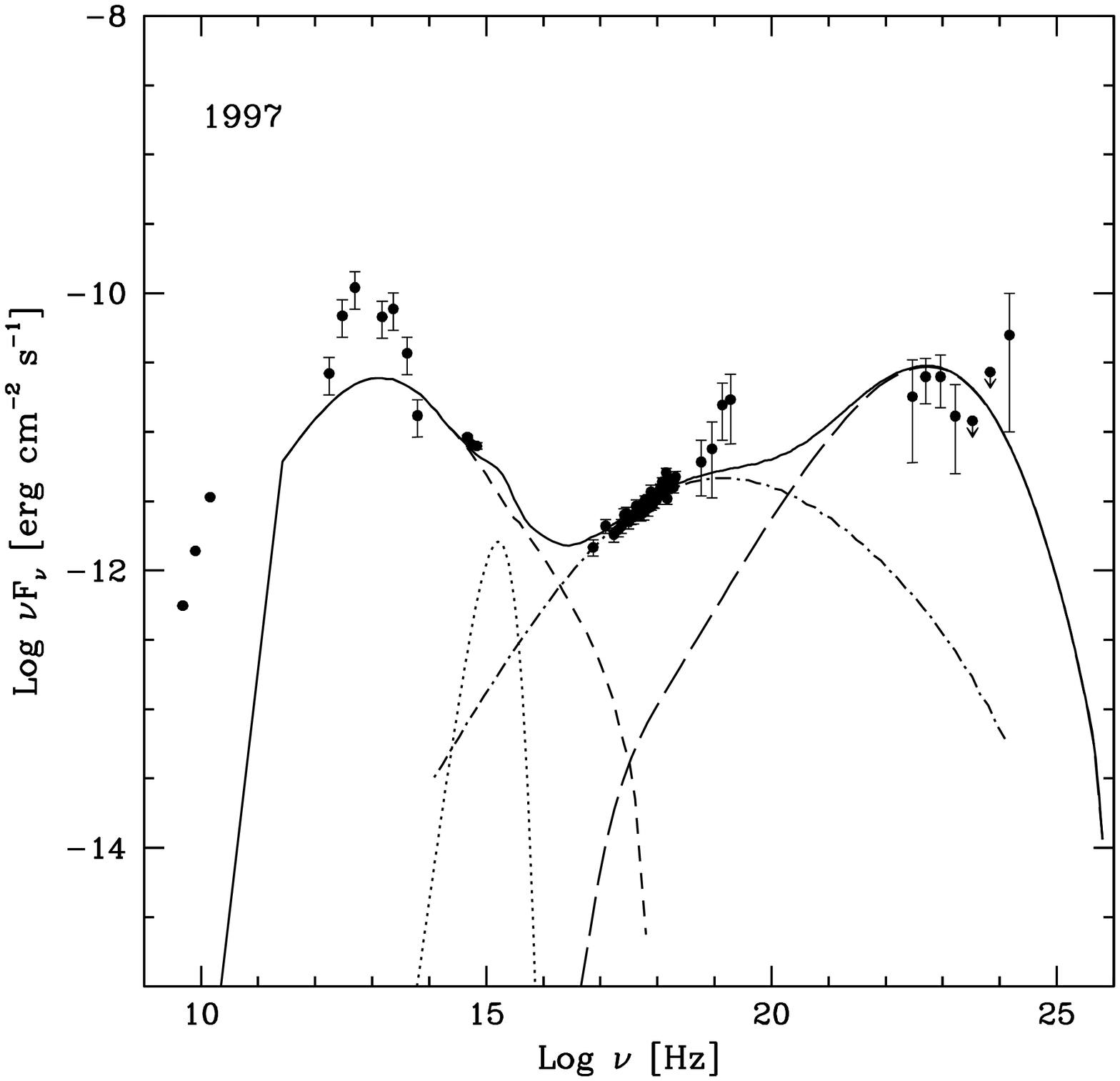]{SED of 3C~279 obtained with simultaneous observations in 1997 January. Short dashed, long dashed, dot-short dashed and dotted curves show the synchrotron, external Compton, synchrotron self-Compton and disk components, respectively (for details of the model see the text). Continuous line is a sum of all the contributions. Points from $10^{12}\,$Hz to $6\cdot10^{13}\,$Hz are ISO data and come from Haas~et~al.~1998; the curve don't fit these data because they are not produced in the region to which we refer in this work (as touched on in the text, they may be emitted by dust external to the jet). \label{fig:sed97} }

\figcaption[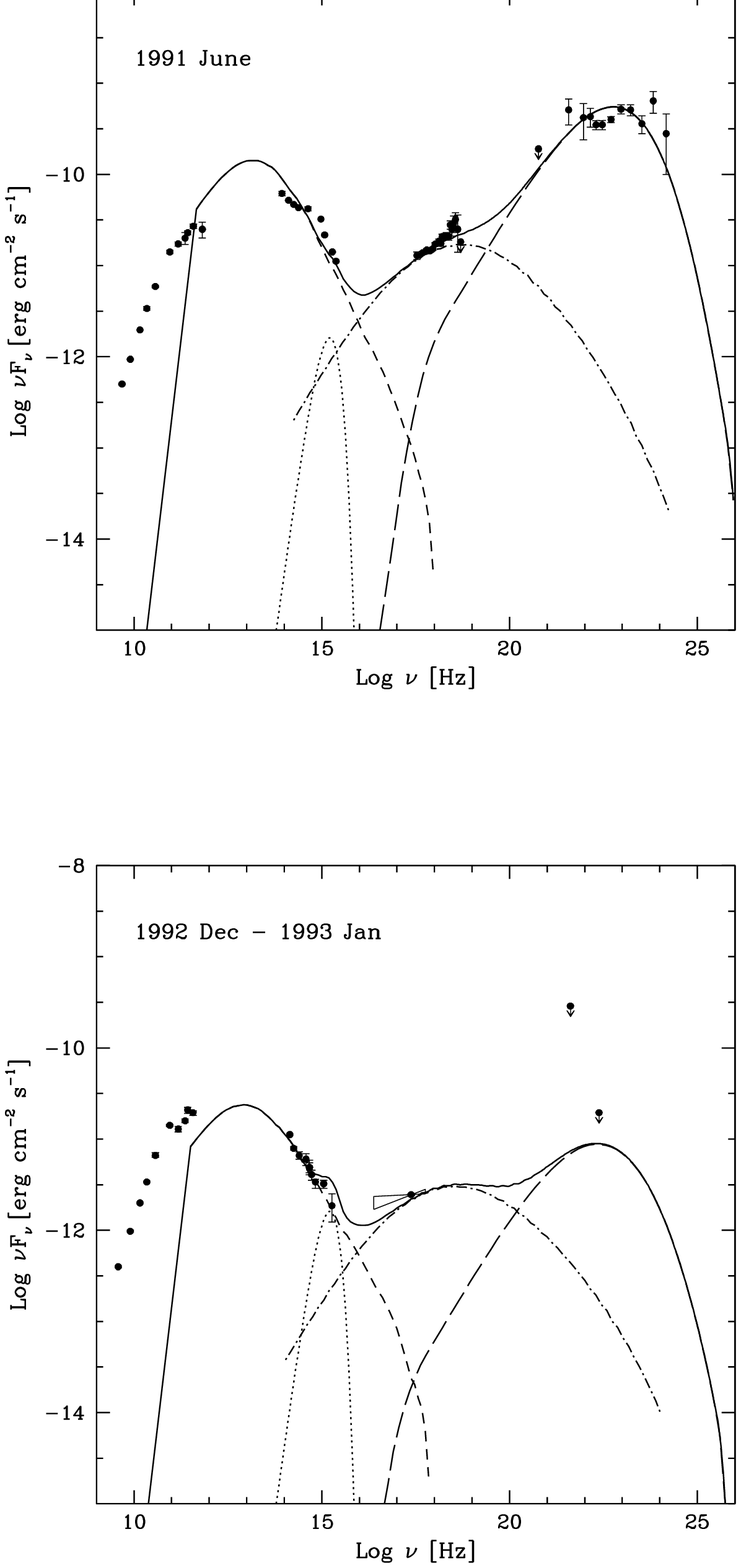]{Overall SEDs of \3 dating back to 1991 June (upper panel) and 1992 December - 1993 January (lower panel). Short dashed, long dashed, dot-short dashed and dotted curves show the synchrotron, external Compton, synchrotron self-Compton and disk components, respectively (for details of the model see the text). Continuous line is a sum of all the contributions. Simultaneous data are from Hartman~et~al.~1996~(1991), Bonnell~et~al.~1994~(1991; UV), Maraschi~et~al.~1994~(1993). \label{fig:sed9193} }

\figcaption[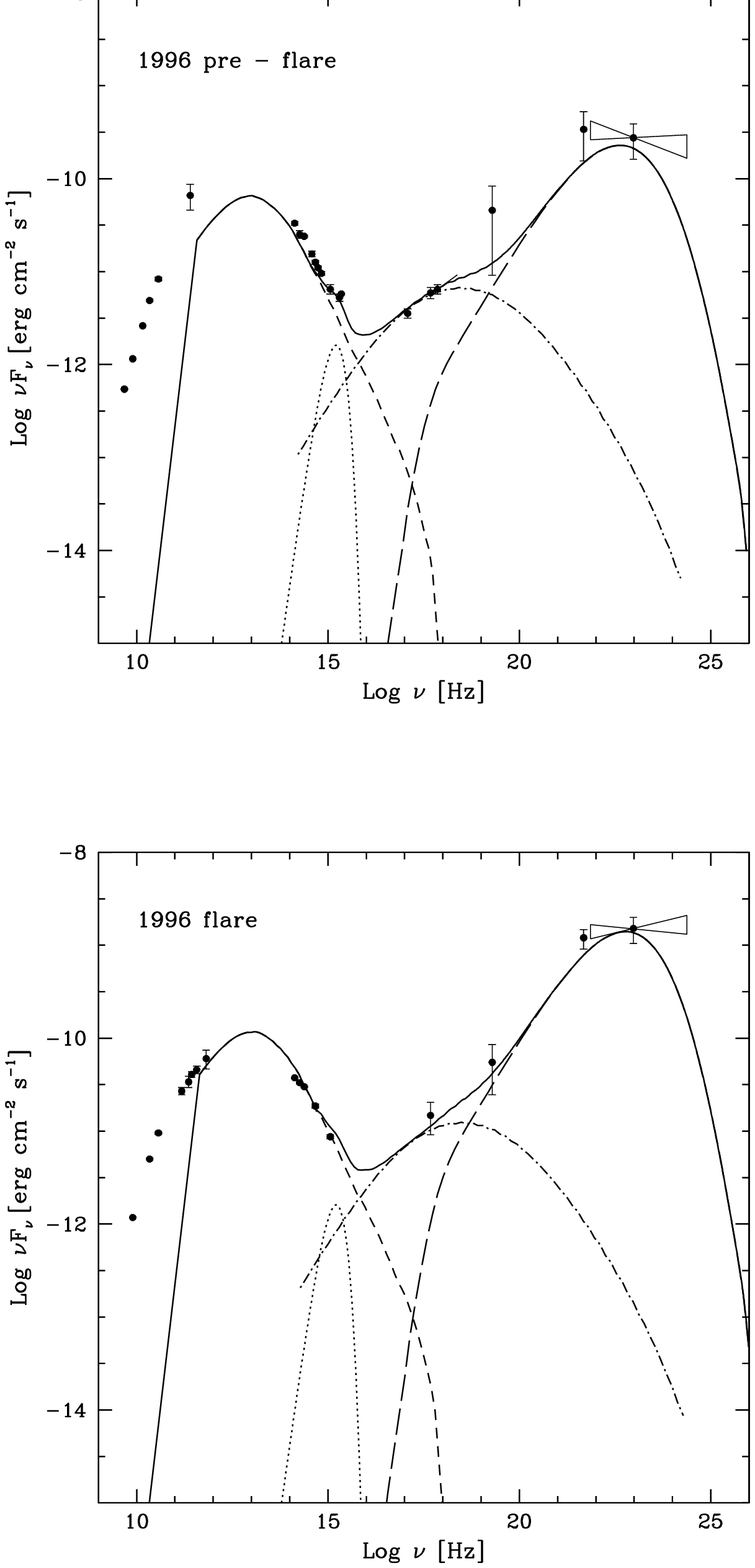]{Overall SEDs of \3 dating back to 1996 January (pre-flare, upper panel) and 1996 February (flare, lower panel). Short dashed, long dashed, dot-short dashed and dotted curves show the synchrotron, external Compton, synchrotron self-Compton and disk components, respectively (for details of the model see the text). Continuous line is a sum of all the contributions. Simultaneous data are from Wehrle~et~al.~1998.\label{fig:sed96}}

\figcaption[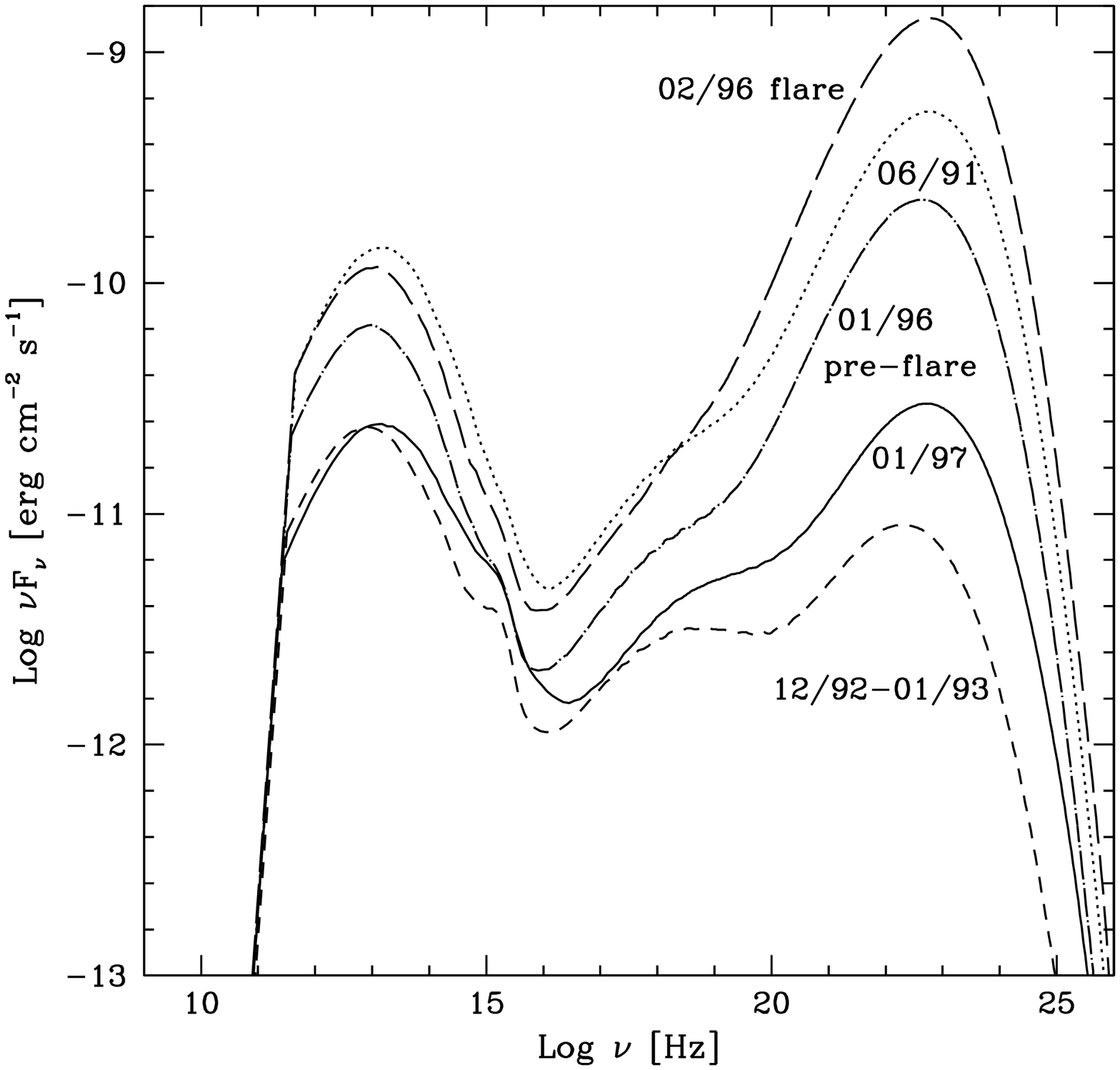]{Comparison between the SEDs of \3 in different levels: the lines reported are sums of all the components used to describe the emission observed (synchrotron radiation, SSC, EC and disk emission). \label{fig:cfr} }

\figcaption[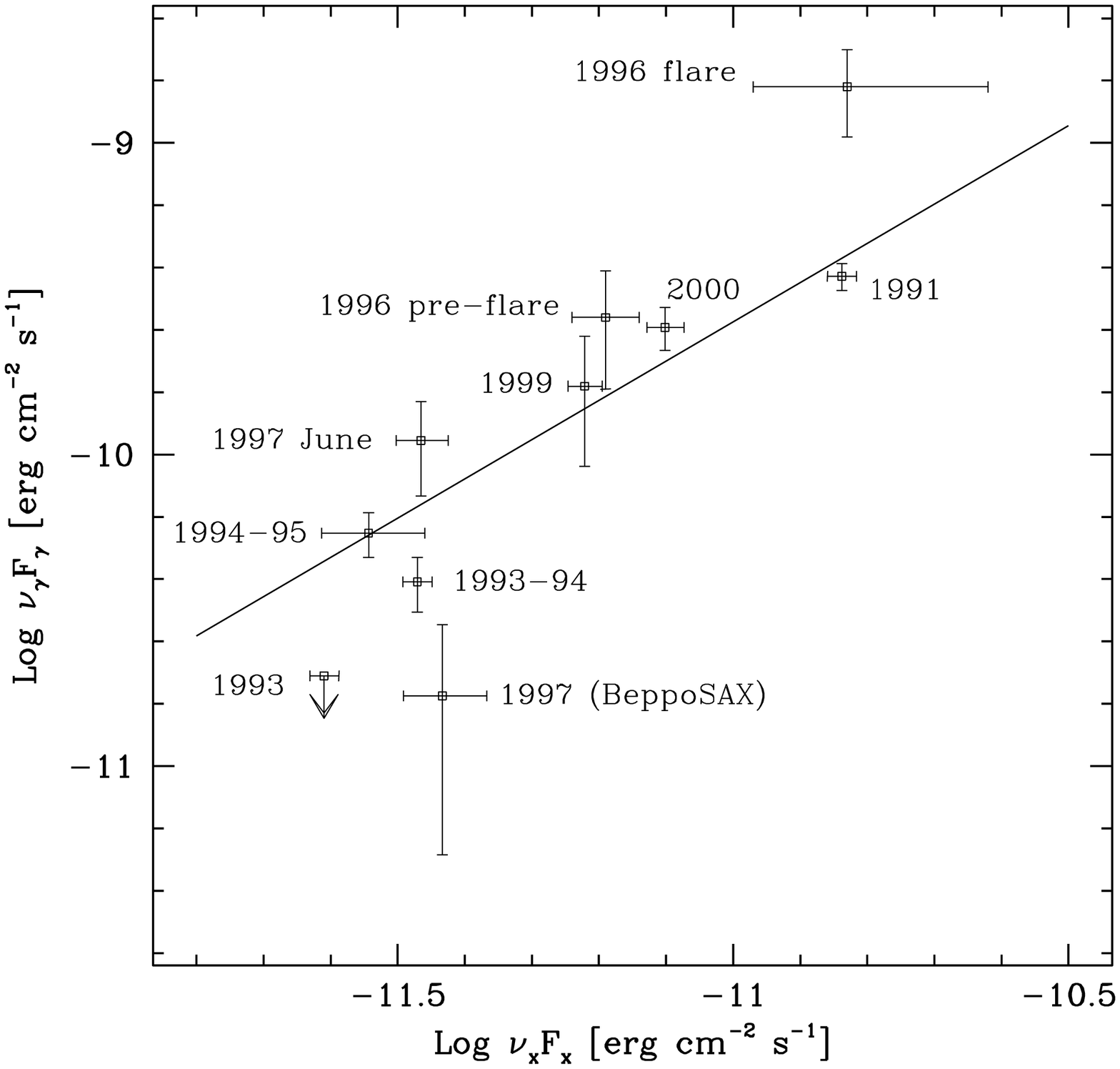]{$\gamma$-ray~vs~X-ray flux for different source states. To the former we used fluxes at $167\,$MeV, while the latter has been measured at $3\,$keV. For several states the we have only a maximum and a minimum flux in the X-ray band. Data are from H01 (see this article for the references); we thought right to use as errors in both the ranges the maximum between the values provided and the $5\%$ of the flux, in order to take into accounts intercalibrations problems. The fit with a straight line provides a slope $1.49\pm0.13$, with $\chi^2_r=2.76$. \label{fig:correl} }

\begin{landscape}
 \begin{table*}
  \begin{center}
  \caption{\sax data observation log.}\label{tab:saxlog}
  \vspace{1cm}
 %\hspace*{-1.5cm}
   \begin{tabular}{r @{/} c @{/} l r @{:} c @{:} l r @{:} c @{:} l c r @{.} l @{$\pm$} r @{.} l c r @{.} l @{$\pm$} r @{.} l c r @{.} l @{$\pm$} r @{.} l }
    \hline \hline
    \multicolumn{3}{c}{} & \multicolumn{3}{c}{} & \multicolumn{3}{c}{} & \multicolumn{4}{c}{LECS$\,^{(a)}$} &  & \multicolumn{4}{c}{MECS$\,^{(b)}$} &  & \multicolumn{4}{c}{PDS$\,^{(c)}$} &  \\
    \multicolumn{3}{c}{Date} & \multicolumn{3}{c}{Start} & \multicolumn{3}{c}{End} & Exp. & \multicolumn{4}{c}{net cts/s} & Exp. & \multicolumn{4}{c}{net cts/s} & Exp. & \multicolumn{4}{c}{net cts/s} \\
    \multicolumn{3}{c}{} & \multicolumn{3}{c}{} & \multicolumn{3}{c}{} & ($10^4\,$s) & \multicolumn{4}{c}{($10^{-1}\,$)} & ($10^4\,$s) & \multicolumn{4}{c}{($10^{-1}$)} & ($10^4\,$s) & \multicolumn{4}{c}{($10^{-1}$)} 
    \vspace{0.1cm}\\
    \hline
    \multicolumn{24}{c}{}\\
    13-23 & 01 & 97 & 22 & 55 & 25 & 23 & 13 & 06 & 3.116 & 0 & 67 & 0 & 02 & 8.473 & 1 & 14 & 0 & 01 & 3.653 & 1 & 11 & 0 & 46
    \vspace{0.5cm} \\
    \hline
    \multicolumn{24}{c}{}\\
    \multicolumn{24}{l}{{\footnotesize$^{(a)}\,$0.1-4.5 keV}}\\
    \multicolumn{24}{l}{{\footnotesize$^{(b)}\,$1.8-10.5 keV, 3 MECS units}}\\
    \multicolumn{24}{l}{{\footnotesize$^{(c)}\,$15-180 keV}}\\
   \end{tabular}
  \end{center}
 \end{table*}
\end{landscape}

{\scriptsize
\begin{landscape}
 \begin{table*}
  \begin{center}
  \caption{Fits to \sax Data (LECS+MECS+PDS): first line simple power-law, second line broken power-law, third line power-law+black body model. Errors are quoted at the 90\% confidence level for 1 parameter of interest ($\Delta \chi ^2=2.71$).}\label{tab:saxfit}
  \vspace{1cm}
   \begin{tabular}{rccccccl}
    \hline
    \hline
    \multicolumn{1}{c}{model} & $\Gamma\,^{(a)}$ & $\Gamma_s\,^{(b)}$ & $E_b\,^{(b)}$ & $KT\,^{(c)}$ & $N_H$                     & $F_{[2\,-\,10\,\rm keV]}$                 & \multicolumn{1}{c}{$\chi^2/$d.o.f.}\\
                              &                  &                    &     (keV)     &     (keV)    &($\times10^{20}$ cm$^{-2}$)&($\times10^{-12}\,$erg cm$^{-2}\,$s$^{-1}$)& 
    \vspace{0.2cm}\\
    \hline
    \multicolumn{8}{c}{}\\
    simple power-law & $1.66\pm0.04$ & -- & -- & -- & $2.10^{+0.52}_{-0.49}$ & 5.79 & 114.9/123 \\
    \multicolumn{8}{c}{}\\
    broken power-law & $1.66\pm0.05$ & $2.40^{+1.22}_{-0.64}$ & $0.82^{+3.10}_{-0.26}$ & -- & $3.98^{+2.86}_{-1.68}$ & 5.80 & 111.2/121 (86$\%$)\\
    \multicolumn{8}{c}{}\\
    power-law+black body & $1.65\pm0.05$ & -- & -- & $0.11\pm0.03$ & $3.63^{+2.90}_{-1.55}$ & 5.81 & 111.4/121 (85$\%$)
    \vspace{0.2cm}\\
    \hline
    \multicolumn{8}{c}{}\\
    \multicolumn{8}{l}{$^{(a)}$ photon index, related to the spectral index $\alpha$ (where $F\propto\nu^{-\alpha}$) by $\alpha =\Gamma-1$.} \\
    \multicolumn{8}{l}{$^{(b)}$ break energy and high energy photon index for the broken power-law model.} \\
    \multicolumn{8}{l}{$^{(c)}$ temperature for the black body model.} \\
   \end{tabular}
  \end{center}
 \end{table*}
\end{landscape}
}

%\begin{landscape}
 \begin{table*}
  \begin{center}
  \caption{Optical observation log.}\label{tab:optlog}
  \vspace{1cm}
 %\hspace*{-1.5cm}
   \begin{tabular}{r @{/} c @{/} l r @{.} l @{$\pm$} r @{.} l r @{.} l @{$\pm$} r @{.} l r @{.} l @{$\pm$} r @{.} l }
    \hline \hline
    \multicolumn{3}{c}{Date} & \multicolumn{12}{c}{Magnitude} \\
    \multicolumn{3}{c}{} & \multicolumn{4}{c}{R} & \multicolumn{4}{c}{V} & \multicolumn{4}{c}{B} \\
    \multicolumn{3}{c}{} & \multicolumn{4}{c}{($4.36\cdot10^{14}\,$Hz)} & \multicolumn{4}{c}{($5.37\cdot10^{14}\,$Hz)} & \multicolumn{4}{c}{($6.76\cdot10^{14}\,$Hz)}
    \vspace{0.1cm}\\
    \hline
    \multicolumn{15}{c}{}\\
    11 & 01 & 97 & 15 & 50 & 0 & 03 & 15 & 93 & 0 & 04 & \multicolumn{4}{c}{--} \\
    12 & 01 & 97 & 15 & 46 & 0 & 02 & 15 & 88 & 0 & 03 & 16 & 31 & 0 & 07 \\
    13 & 01 & 97 & 15 & 37 & 0 & 03 & 15 & 84 & 0 & 04 & 16 & 37 & 0 & 05 \\
    14 & 01 & 97 & 15 & 50 & 0 & 02 & \multicolumn{4}{c}{--} & \multicolumn{4}{c}{--} \\
    15 & 01 & 97 & 15 & 48 & 0 & 04 & 15 & 99 & 0 & 05 & \multicolumn{4}{c}{--} \\
    16 & 01 & 97 & 15 & 48 & 0 & 02 & 15 & 95 & 0 & 07 & \multicolumn{4}{c}{--} \\
    17 & 01 & 97 & 15 & 49 & 0 & 02 & 15 & 98 & 0 & 03 & \multicolumn{4}{c}{--} \\
    18 & 01 & 97 & 15 & 52 & 0 & 02 & 16 & 01 & 0 & 04 & 16 & 55 & 0 & 05 \\
    \multicolumn{3}{c}{}  & 15 & 57 & 0 & 02 & 15 & 98 & 0 & 03 & \multicolumn{4}{c}{--} \\
    \multicolumn{3}{c}{}  & 15 & 50 & 0 & 03 & 16 & 07 & 0 & 04 & \multicolumn{4}{c}{--} \\
    \multicolumn{3}{c}{}  & 15 & 50 & 0 & 04 & \multicolumn{4}{c}{--} & \multicolumn{4}{c}{--}
    \vspace{0.5cm} \\
    \hline
   \end{tabular}
  \end{center}
 \end{table*}
%\end{landscape}

\begin{table*}
 \begin{center}
  \caption{Parameters used for the emission model described in the text; the BLR luminosity is $L_{blr}=6\cdot10^{44}\,$erg~s$^{-1}$. We assume a blob with $R=5\cdot10^{16}\,$cm in this radiation field diluted in a region with radius $R_{blr}=4\cdot10^{18}\,$cm. The electron distribution extended from $\gamma_{min}=1$ to $\gamma_{max}=5\cdot10^{4}$, and the angle of view is $\theta=3^{\circ}$. These values was kept fixed in all fits (see the text for more details).}\label{tab:parameters}
  \vspace{1cm}
  \begin{tabular}{cccccccccccc}
   \hline
   \hline
   $\Gamma$ & \quad\quad\quad & $\delta^{(a)}$ & \quad\quad\quad & $B$ &  \quad\quad\quad &  $\gamma_b$  &  \quad\quad\quad & $n_1$ & \quad\quad\quad & $n_2$ &  $k$ \\
            &            &          &            &($G$)&        &($\times10^2$)&        &       &            &       &($\times10^3\,$cm$^{-3}$)
   \vspace{0.1cm} \\
   \hline
   \multicolumn{12}{c}{}\\
   \multicolumn{12}{c}{\bf 1991}
   \vspace{0.1cm} \\
   \hline
   13 & & 17.8 & & 0.6 & & 5.5 & & 1.6 & & 4.7 & 5
   \vspace{0.5cm} \\
   \multicolumn{12}{c}{\bf 1993}
   \vspace{0.1cm} \\
   \hline
   6 & & 10.9 & & 0.7 & & 4.5 & & 1.6 & & 4.4 & 5.6
   \vspace{0.5cm} \\
   \multicolumn{12}{c}{\bf 1996 pre~-~flare}
   \vspace{0.1cm} \\
   \hline
   11 & & 16.5 & & 0.5 & & 5 & & 1.6 & & 4.7 & 5
   \vspace{0.5cm} \\
   \multicolumn{12}{c}{\bf 1996 flare}
   \vspace{0.1cm} \\
   \hline
   17 & & 19 & & 0.5 & & 4.9 & & 1.6 & & 4.7 & 5.3
   \vspace{0.5cm} \\
   \multicolumn{12}{c}{\bf 1997}
   \vspace{0.1cm} \\
   \hline
   7 & & 12.3 & & 0.5 & & 6.0 & & 1.6 & & 4.2 & 4.5  
   \vspace{0.5cm} \\
   \hline
   \multicolumn{12}{c}{}\\
   \multicolumn{12}{l}{$^{(a)}$ completely determined if $\Gamma$ and $\theta$ are fixed; therefore this is not a free parameter}
  \end{tabular}
 \end{center}
\end{table*}

\clearpage

\begin{figure}
 \centerline{\plotone{f1.ps}}
\end{figure}

\clearpage

\begin{figure}
 \centerline{\plotone{f2.ps}}
\end{figure}

\clearpage

\begin{figure}
 \centerline{\plotone{f3.ps}}
\end{figure}

\clearpage

\begin{figure}
 \centerline{\plotone{f4.ps}}
\end{figure}

\clearpage

\begin{figure}
 \centerline{\plotone{f5.ps}}
\end{figure}

\clearpage

\begin{figure}
 \centerline{\plotone{f6.ps}}
\end{figure}

\end{document}